\documentclass[12pt,preprint]{aastex}

\usepackage{graphicx}
\usepackage{subfigure}
\usepackage{amsmath}
\usepackage{amssymb}

\shorttitle{SDO high temperature strands in coronal active regions}
\shortauthors{Reale et al.}

\begin{document}

\title{Solar Dynamics Observatory discovers thin high temperature strands in coronal active regions}

\author{Fabio Reale\altaffilmark{1,2}, Massimiliano Guarrasi\altaffilmark{1}, Paola Testa\altaffilmark{3}, Edward E. DeLuca\altaffilmark{3}, Giovanni Peres\altaffilmark{1,2}, Leon Golub\altaffilmark{3}}

% \affil{Dipartimento di Fisica, Universit\`a di Palermo, Piazza del Parlamento 1, 90134 Palermo, Italy}
% \affil{Harvard-Smithsonian Center for Astrophysics, Cambridge, MA 02138, USA}
\altaffiltext{1}{Dipartimento di Fisica, Universit\`a di Palermo, Piazza del Parlamento 1, 90134 Palermo, Italy}
\altaffiltext{2}{INAF - Osservatorio Astronomico di Palermo ``G.S. Vaiana'', Piazza del Parlamento 1, 90134 Palermo, Italy}
\altaffiltext{3}{Harvard-Smithsonian Center for Astrophysics, Cambridge, MA 02138, USA}

\begin{abstract}
One scenario proposed to explain the million degrees solar corona is a finely-stranded corona where each strand is heated by a rapid pulse. However, such fine structure has neither been resolved through direct imaging observations nor conclusively shown through indirect observations of extended superhot plasma.
Recently it has been shown that the observed difference in appearance of cool and warm coronal loops ($\sim1$~MK, $\sim2-3$~MK, respectively) -- warm loops appearing "fuzzier" than cool loops -- can be explained by models of loops composed of subarcsecond strands, which are impulsively heated up to $\sim10$~MK. That work predicts that images of hot coronal loops ($\gtrsim6$~MK) should again show fine structure.
Here we show that the predicted effect is indeed widely observed in an active region with the Solar Dynamics Observatory, thus supporting a scenario where impulsive heating of fine loop strands plays an important role in powering the active corona.
\end{abstract}


\section{Introduction}

The bright corona consists of magnetic loop-like tubes which confine the heated plasma.
It has been proposed that the plasma is heated by rapid energy pulses, the so-called nanoflares, due to very localized reconnections of the magnetic field braided and twisted by the chaotic motions of the loop footpoints in the photosphere \citep{parker88,cargill94,cargill04}. Although impulsive events were predicted to be small and rapid, they are expected to produce hot, variable emission at $\sim10$ MK, even in the absence of flares. The evidence for these signatures has been neither direct nor conclusive so far \citep{reale09,reale09b,mctiernan09,schmelz09,sylwester010}. The lack of evidence suggests that alternative mechanisms, i.e. the more gradual dissipation of MHD Alfven waves \citep{hollweg84,nakariakov99,ofman08}, or even the direct involvement of the underlying chromosphere \citep{2011Sci...331...55D}, are important (even critical) to the solution of the heating problem.

Indeed, there are good reasons why extensive nanoflaring activity has been elusive so far \citep{klimchuk06,reale010}. An important one is that, since  momentum and energy are transported only along the magnetic field lines, a single pulse heats just one long and thin strand at a time. These strands are too thin to be spatially resolved by current telescopes, and the detection of hot plasma is difficult because of its small filling factor \citep{1985SoPh...96..253M}.

Recently, it has been shown that coronal loops substructured in a multitude of thin strands pulse-heated up to 10~MK can explain the evidence of increasing fuzziness with temperature of emitting plasma \citep{guarrasi010}. This model also predicts that the fuzziness will decrease in bands sensitive to temperature $>3$~MK.  
Here we show that this prediction is indeed confirmed by active region observations with the Solar Dynamics Observatory.

We suppose that, in the multi-stranded loops that populate an active region, the duration of each energy pulse is much smaller than the plasma characteristic radiative and conductive cooling times \citep{serio91}. Once the heat pulse ends, the plasma cools down exponentially. If each loop strand is heated infrequently, i.e. the time between heating events is comparable or larger than the radiative cooling time, the plasma in the strand will be at high temperature only for a short time, and cooling for most of the time. Therefore, at a given time, only a few strands will be very hot, and many of them will be significantly cooler than their peak temperature. If nanoflares are scaled-down versions of coronal flares, we expect peak temperatures of about 10~MK. On the other hand, while the plasma is so hot, its emission is faint because of its low density: 
dense (and bright) plasma comes up from the chromosphere on time scales longer than
the duration of the heat pulses. Therefore, the strands become bright when the plasma is already cooling and remain bright until the plasma has drained significantly. More specifically, it has been shown that a strand heated to 10~MK for $\sim1$~min is at its brightest when it has cooled to $\sim3$~MK, and that it remains around this temperature for a relatively long time \citep{guarrasi010}. Thus, in this scenario, we expect to see few hot 10~MK strands, and many 3~MK strands in active regions. 

\section{Data analysis and modeling}

The Solar Dynamics Observatory (SDO) was launched in February~2010.
% and was designed to provide continuously spectra and images of the Sun in several 
% wavebands and at very high cadence. 
The Atmospheric Imaging Assembly (AIA), with its 7 EUV narrowband
channels, images the solar corona with high spatial resolution ($\sim0.6$arcsec/pixel), and high cadence in several spectral bands at the
same time.
We analyze SDO/AIA observations of AR11117 on 28~October~2010, from 2:00~UT. We consider a $500\times500$~pixels region in three channels: 171\AA, 335\AA, 94\AA.
The narrow passbands of the AIA channels contain bright spectral lines emitted by plasma at different temperatures. In particular, the channels centered at 94\AA\  and 335\AA\ contain a strong line of FeXVIII and FeXVI, respectively, which are emitted more efficiently by plasma around 8~MK and 3~MK \citep{odwyer010}. 
Therefore, these two channels are appropriate to test the prediction of the model.

\begin{figure}[htbp]               %%%%%%Figura%%%%%%%%%
 \centering
   {\includegraphics[width=10cm]{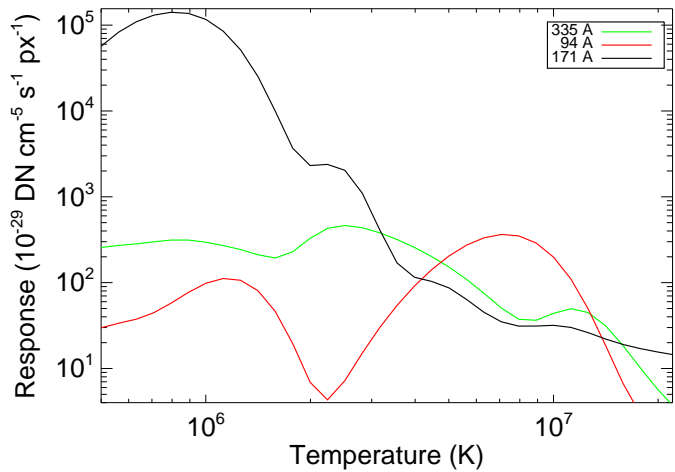}}
\caption{\small Instrument response per unit emission measure and as a function of temperature, for the 94\AA,  335\AA\ and 171\AA\ SDO/AIA channels.}
\label{fig:aia_gt}
\end{figure}

\begin{figure}
  \centering
%   \flushleft{\large \bf ~a\hspace{7.7cm}b}
  \includegraphics[width=9cm]{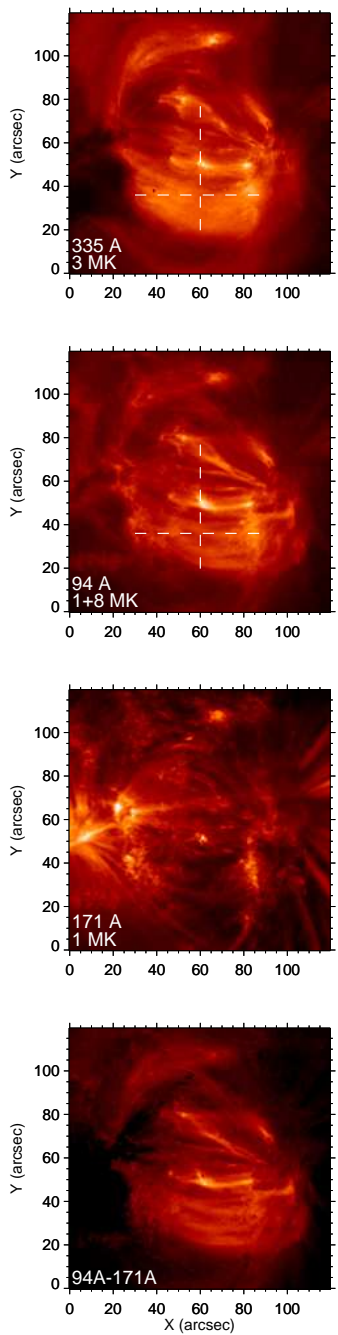}
\caption{\small Inner part of the active region AR11117 observed in 3 different channels (335\AA, 94\AA, 171\AA) of the SDO/AIA on 27~October~2010 around 02~UT. The color scales as the square root of the pixel counts. The ranges are 59--1567, 55--3157, 155--4681 DN for the three channels, respectively. The channels are most sensitive to plasma emitting at the labeled temperatures. The bottom panel shows the image obtained by subtracting the cool component scaled from the 171\AA\ channel to the 94\AA\ image (same color scale as second panel, real range 0--2741 DN). Fig.~\ref{fig:rms} shows brightness profiles along the marked vertical and horizontal lines.}
\label{fig:ar}
\end{figure}

We use level-1.0 data, after standard processing of level-0 data (bad-pixel removal, despiking, flat-fielding). Data are obtained from a standard observing series with cadence of 12~s in all channels, and exposure times of 2~s in the 171\AA\
channel and 2.9~s in the 335\AA\ and 94\AA~channels. The channels have very different sensitivity to the solar coronal emission (Fig.~\ref{fig:aia_gt}), resulting in different signal-to-noise ratios.
Therefore, for a more meaningful comparison of the active region morphology in the different channels, and to avoid possible
spurious effects due to the different noise level, we summed
images in the lower intensity channels: first we coaligned the images in each channel by using a
standard cross-correlation routine (tr\_get\_disp.pro in IDL SolarSoftware package); then we added 30 consecutive images in the 94\AA~channel, and
3 in the 335\AA\ channel (noise of $\sim3-4$\% in both channels in the central region of Fig.~\ref{fig:ar}).
Final images in all three channels were then co-aligned again.
We note that during the
94\AA\ time interval ($\approx7$~min) the region variability is of a
few percent at most, and that our conclusions on the observed morphological differences in the hotter vs.\ cooler emission are, if anything, underestimated
by the temporal averaging we effectively apply by summing up images.

% Fig.~\ref{fig:fov} shows the whole active region. 
% We notice that in the 335\AA\ channel, the core, discussed in detail in the main text, is surrounded by a halo of fainter and larger arches (marked with LL), which concentrically depart from two poles located at the east and west of the core of the active region. In the 94\AA\ channel, the large arches around the core are fainter and hardly visible, making the appearance of the region less diffuse. In the 171\AA\ channel, giant arches (marked with OL) depart from the two magnetic poles, but mostly in the outbound east-west direction. The bright moss extends from the core also to the north-east.

Figure~\ref{fig:ar} (top three images) shows the internal part of the active region in the 94\AA, 335\AA\  and 171\AA\  channels. The 171\AA\ channel contains a very strong FeIX line with peak emissivity at $\approx0.8$~MK.
% mostly emitted by plasma around 1 MK. 
For the moment, we focus our attention to the top two images, and, in particular, to the core of the active region. 

In the 335\AA\ channel, the active region core is covered quite uniformly by a large number of bright arches. In the southern part ($20\lesssim Y\lesssim 55$ arcsec), the arches coalesce to form a uniform bright band. In the northern part, we can identify three brighter loop groups in a background of more diffuse emission. A few very bright spots are visible at the center of the region. Overall the region has quite a diffuse appearance and individual loops cannot be clearly resolved. 

In the 94\AA\ channel, overall we see a very similar morphology and many bright structures are clearly cospatial with those observed in the 335\AA\ channel. The most striking difference from the image in the other channel is in the core itself: in the southern part, while in 335\AA\ the arches are densely packed and uniform, in 94\AA\ they have greater contrast, i.e. we see an alternation of bright and fainter structures.
In the northern part, we are even able to resolve very thin bright east-to-west bridges, in the same location where thicker arches are present in the softer channel. Overall, in the 94\AA\  channel, the loop systems appear sharper, the observed emission largely less ``fuzzy'', and we can resolve thinner bright structures than in the 335\AA\  channel.
This is exactly the effect that we expected, and that was predicted \citep{guarrasi010}. 

We have however to be cautious in one important point.
Although the channel passbands are narrow, they include several spectral lines. In particular, the 94\AA\ channel includes another strong line (FeX) which peaks at $\sim1$~MK. In general, we cannot be sure that the emission imaged by this channel comes only from hot plasma ($\gtrsim6$ MK). The 171\AA\  image helps us in this respect, because it allows us to localize the bright cooler plasma ($T\sim1$ MK), and to assess whether the 94\AA\  emission is due to hot or cool plasma. In the 171\AA\ channel the active region shows quite a different morphology. Many structures are complementary to those observed in the other channels \citep{reale07}. The core appears depleted of arch-like structures. Only few of them are visible, and they look quite different from those in both the other channels. The arch-like structures are instead replaced by bright ``moss". This moss is a well-known feature of this soft channel, already studied in {\it Normal Incidence X-ray Telescope} (NIXT) and {\it Transition Region and Coronal Explorer} (TRACE) observations and commonly explained as the bright warm footpoints of (hot) high-pressure loops \citep{peres94,fletcher99,martens00}. The 171\AA\  image clearly indicates that much of the plasma confined in the filamented arches that we see in the 94\AA\  channel is not warm at 1 MK, and therefore it must be hot around 6--8~MK.  As a further test, we estimated the emission measure of the cool plasma from the 171\AA\ map, and used it to compute the expected contribution of the cool plasma in the 94\AA\ channel. We made the conservative assumptions that the cool plasma temperature is the one of the peak of the cool 94\AA\ response component (Fig.~\ref{fig:aia_gt} ), i.e. $\log T=6.05$, and that the cool 94\AA\ response component is underestimated by a factor 5 \citep{markus011}. We subtracted this estimated cool 94\AA\ map from the 94\AA\ map shown in Fig.~\ref{fig:ar}. The resulting image is also shown in Fig.~\ref{fig:ar}(bottom). We note that the moss emission visible in the 94\AA\ channel is considerably reduced, and that the loop footpoints regions and the outbound loops are mostly removed. On the other hand, the emission of the loop bundle in the central region does not change significantly, except for a reduction in the moss region, and  the fine structures in the upper half of the image are also almost unchanged. This confirms that the emission of the loop bundle and of the fine structures mostly comes from hot plasma.

\begin{figure}
  \centering
%   \flushleft{\large \bf ~a\hspace{7.7cm}b}
 \subfigure[]
  {\includegraphics[width=8cm]{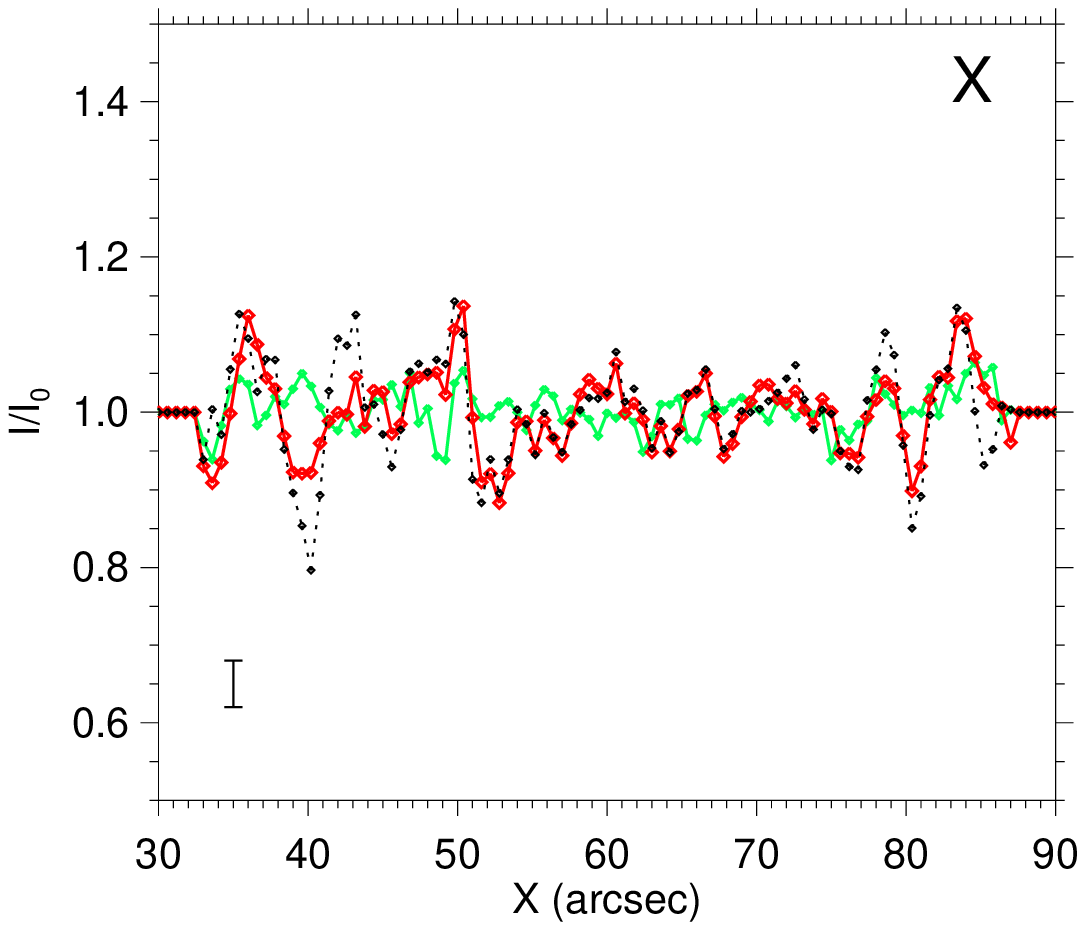}}
 \subfigure[]
  {\includegraphics[width=8cm]{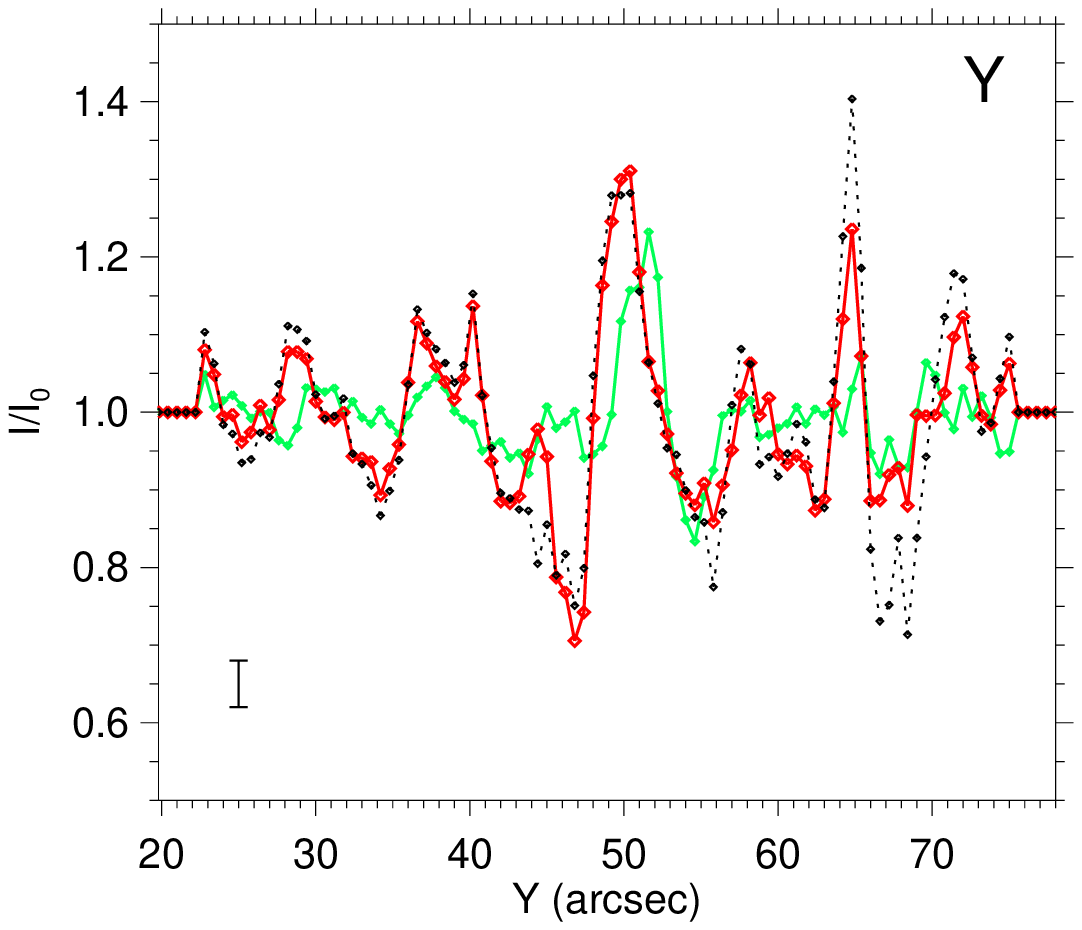}}
  \subfigure[]
  {\includegraphics[width=8.cm]{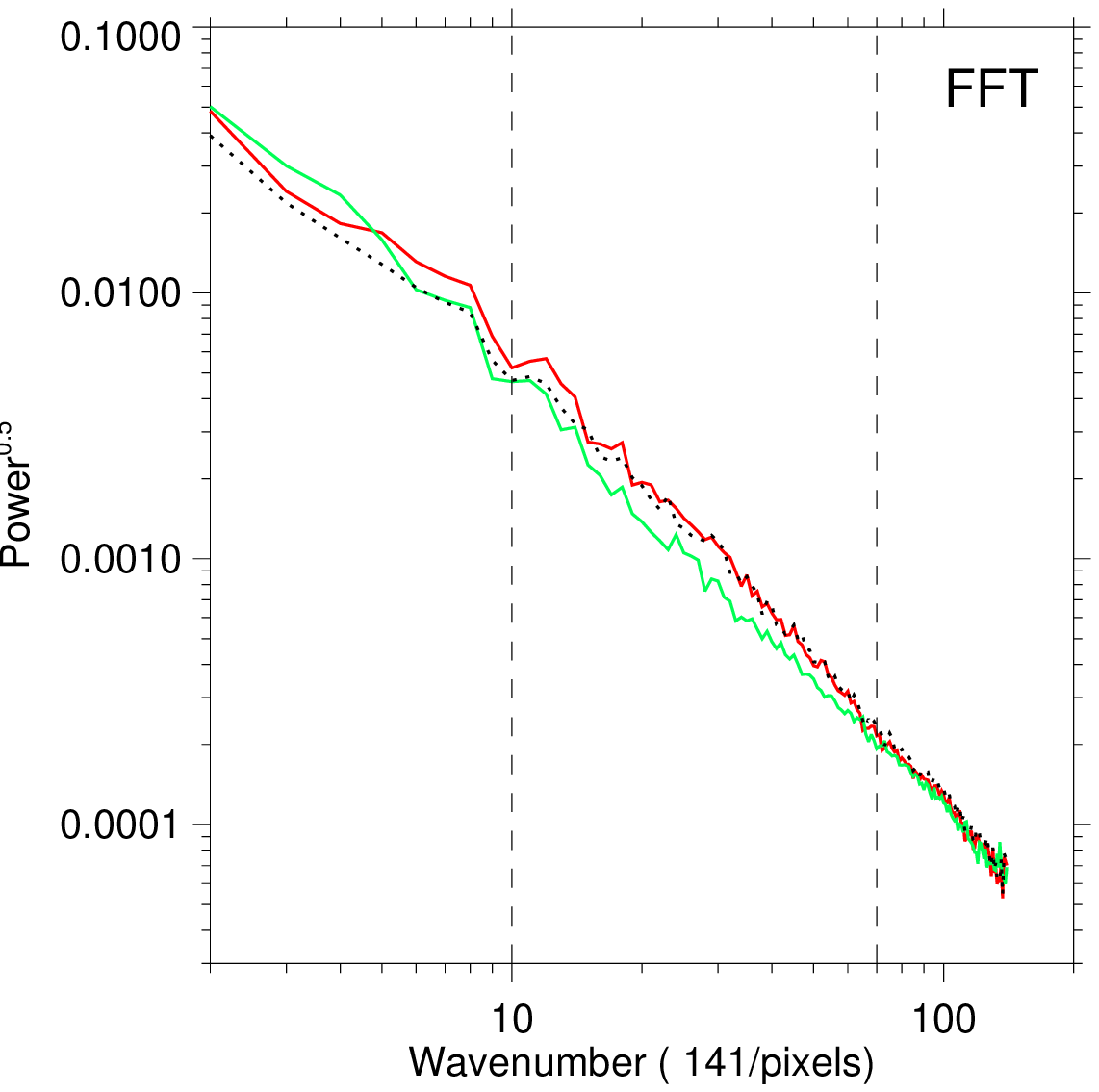}}
\caption{\small Brightness profiles in the 94\AA\, before (red) and after (black dotted) subtracting the cool 1~MK component, and in the 335\AA\ (green) channels along the horizontal (a) and vertical (b) lines in Fig.~\ref{fig:ar}. The profiles are normalized to a moving average with a 10 pixels boxcar. A typical error bar is also shown. (c) Normalized 2-D Fourier Transforms of 335\AA\ and both 94\AA\ images in Fig.~\ref{fig:ar}. 94\AA\ transforms are both systematically higher than the 335\AA\ transform in the wavenumber range between the dashed lines.}
\label{fig:rms}
\end{figure}

For a quantitative estimate of the different fuzziness of the hot ($\gtrsim6$~MK) and cooler ($\sim3$~MK) plasma, Figure~\ref{fig:rms}a,b plots the pixel brightness, divided by a moving average, along the vertical and horizontal lines in Figure~\ref{fig:ar}. The brightness of the 94\AA\ channel after subtracting the cool 1~MK component is also shown.  We note that all the steep gradients are traced by more than one data point, i.e. the trends are coherent and are due to real structures not to noise.
% certifies the evidence of filling factor of emitting plasma changing from 6 - 8 MK to 3 MK more quantitatively. 
Along the horizontal line -- that runs approximately along the magnetic tubes -- the pixel brightness changes in a similar way and with a similar amplitude in both channels. Table~\ref{tab:rms} shows the root-mean-square (RMS) averages of the amplitude excursion, i.e. $\approx3$\% and $\approx5-6$\%, respectively. 
% These might be taken also as upper limits of the photon noise. 
% Along the vertical lines, i.e. across the field line direction, the brightness changes by much more in the 94\AA\  band than in the 335\AA\  band 
Along the vertical line -- which runs {\it across} the field line direction -- both brightnesses in the 94\AA\ channel are significantly more variable in space than in the 335\AA\ channel, and the RMS average of the amplitude excursion becomes 10-13\% vs 6\%.

To check on larger baselines, we derived the analogous values over all the rows ($X$) and columns ($Y$) of selected regions. For each region, Table~\ref{tab:rms} shows the RMS average values, and their standard deviations, in both channels. We selected three regions: one with as little moss as possible, the very center, and a larger core region. For the second region, we also report the values obtained in the 94\AA\ channel after subtracting the cool component. We find that the $Y$-values in the 94\AA\ channel are invariably much higher than all the others, thus confirming that the decrease of fuzziness in the 94\AA\ channel is highly significant.

\begin{deluxetable}{lllllllll}
\tabletypesize{\scriptsize}
% \rotate
\tablecaption{Fractional RMS amplitude excursions\label{tab:rms}}
\tablewidth{0pt}
\tablehead{
\colhead{Region} & \colhead{335\AA\ X}&  \colhead{$\sigma$} & \colhead{94\AA\ X} & \colhead{$\sigma$} & \colhead{335\AA\ Y} & \colhead{$\sigma$} & \colhead{94\AA\ Y} & \colhead{$\sigma$} }
\startdata
Single row/column&0.03& ... &0.05&...&0.06&...&0.10&...\\
Region little moss (size: $60\times60$ pixels, 
&0.032&0.017&0.046&0.021&0.052&0.011&0.082&0.017\\
% Rev 1&  0.032 & 0.017 & 0.059 & 0.019 & 0.052 & 0.011 & 0.094 & 0.017 \\
\mbox{        } center:$ \left[ X=58",Y=38" \right] $)&&&&&&&&\\
% Region no moss (94\AA\-171\AA\) & 0.032 & 0.017 & 0.074 & 0.026 & 0.052 & 0.011 & 0.120 & 0.018 \\
Central core region ($70\times70$, $ \left[ 68",49" \right] $)&
0.044& 0.018&0.059&0.021&0.072&0.016&0.111&0.036\\
% 0.044 & 0.015 & 0.067 & 0.015 & 0.066 & 0.013 & 0.131 & 0.032 \\
Central core region (94\AA--171\AA)&0.044&0.018&0.068&0.025&0.072&0.016&0.130&0.037\\
% Rev 1 0.044 & 0.015 & 0.080 & 0.020 & 0.066 & 0.013 & 0.165 & 0.033 \\
Whole core region ($100\times100$, $\left[ 72",48" \right]$)&0.053&0.021&0.065 &0.021&0.063&0.011&0.099&0.020\\
% Rev1 &  0.053 & 0.021 & 0.076 & 0.018 & 0.063 &  0.011 & 0.115 & 0.025 \\
\enddata
\tablecomments{The first row is computed along the lines marked in Fig.~\ref{fig:ar}. $\sigma$ is the standard deviation of the rms excursions.}
\end{deluxetable}

To add support to this evidence, we take the 2-D Fast Fourier Transforms (FFT) of the images of Fig.~\ref{fig:ar} in the 94\AA\ channel before and after subtracting the cool component and in the 335\AA\ channel. Then we sum along circles of constant wavenumber, as it was done in the past for NIXT observations  \citep{1993ApJ...405..767G,1993ApJ...405..773G}. 
% We probably should apply a Gaussian filter so the signal at the edges of the box go to zero, with will reduce the ringing from applying a periodic FFT to a non-periodic image.
The power at zero wavenumber reflects the mean intensity of the image, so we normalize the images by the average intensity to make cross comparison of the power distributions more straightforward. In Figure~\ref{fig:rms}c the resulting Fourier transforms are power-laws. We see that both 94\AA\ FFTs are systematically higher than the 335\AA\ FFT in the wavenumber range $\sim10-70$, that corresponds to a spatial range $\sim14-2$ pixels. This range is in agreement with the cross-section size of the observed structures. All the lines converge at the highest wavenumbers, which are related to the presence of noise. This indicates a similar signal-to-noise ratio for both channels. In the end, the figure confirms more power at relatively high wavenumbers in the 94\AA\ channel both with and without subtracting the 1~MK component.

% \begin{figure}[htbp]               %%%%%%Figura%%%%%%%%%
%  \centering
%    {\includegraphics[width=10.cm]{f4.ps}}
% \caption{\small Normalized 2-D Fourier Transforms of the images in the 94\AA\ (red) and 335\AA\ (green) channels. It is shown also the transform in the 94\AA\ channel after subtraction of the cool component scaled from the 171\AA\ channel (red dashed).}
% \label{fig:fft2d}
% \end{figure}

\begin{figure}[htbp]
 \centering
 \subfigure[]
   {\includegraphics[width=8cm]{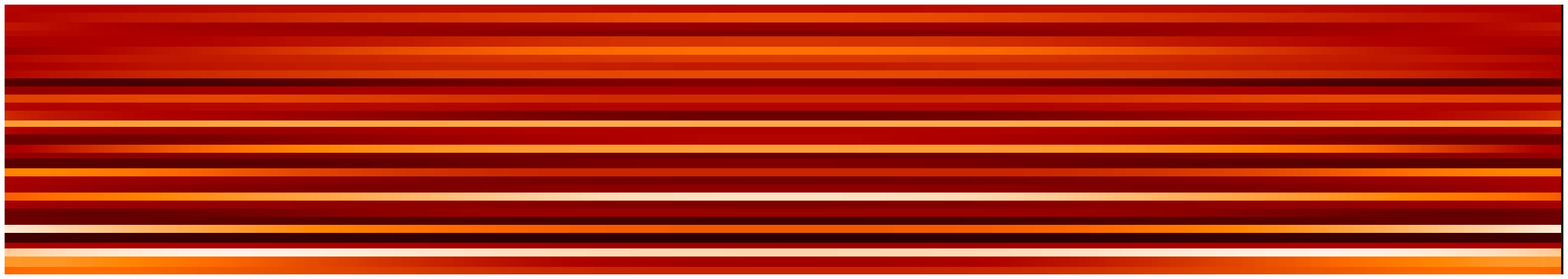}}
   \hspace{10mm}
%    {\includegraphics[width=14.8cm]{335A_lin_dbl_b.ps}}
 \subfigure[]
   {\includegraphics[width=8cm]{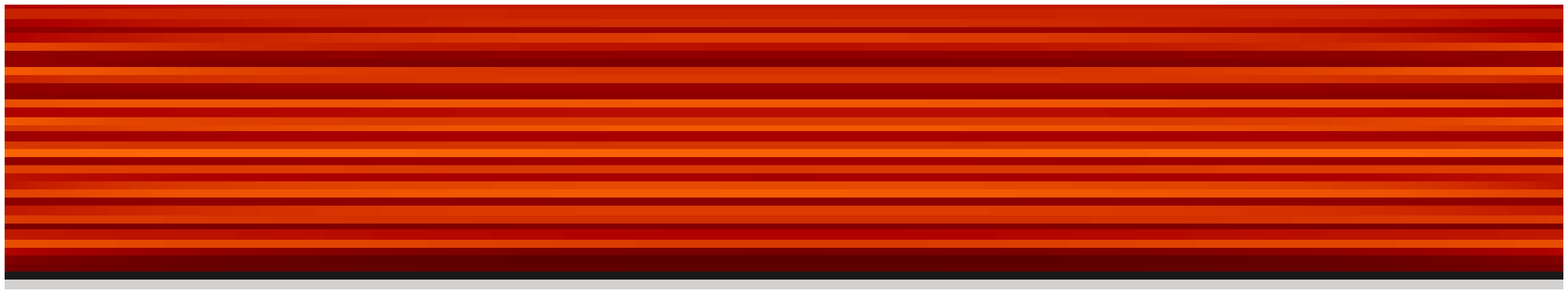}}
% %   \hspace{7mm}
% %    {\includegraphics[width=14.8cm]{CUT_335_95.ps}}  
%  \subfigure[]
%    {\includegraphics[width=9.cm]{FigS5c.ps}}  
\caption{\small Emission of a (straightened) pulse-heated multi-stranded loop synthesized from  hydrodynamic modeling in the (a)  $94$\AA\, and  (b) 335\AA\   channels. The emission in each pixel strip is normalized {\it across} the loop. The color scale is linear between $0.6$ and $1.4$ (average $1$): white is the maximum of emission, and black the minimum. Only the coronal part of the loop (upper 75\% of the loop length) is shown.  
% (c) Cuts across the images (green lines). Red and green lines are related to the SDO $94$\AA\ and the $335$\AA\ channel, respectively.
}
\label{fig:Loops_Emission}
\end{figure}

For reference, we now show some results of hydrodynamic modeling of loops consisting of thin pulse-heated strands \citep{guarrasi010}. Each pulse lasts 60~s and brings the plasma temporarily to $\sim10$~MK. The plasma then cools off freely. Images comparable to the observed ones are obtained, for instance, by assuming that an instrument resolution element includes about 60 unresolved strands. Each strand is symmetric with respect to the loop apex.
Figure~\ref{fig:Loops_Emission} shows the stranded-loop emission obtained by folding the density and temperature profiles from the hydrodynamic model with the 94\AA\, and  335\AA\ channels responses. Only the coronal part of the loop (upper 75\% of the loop length) is shown. To emphasize the effect we are addressing, we normalize the emission in each pixel strip {\it across} the loop. 
In Fig.~\ref{fig:Loops_Emission} we can clearly see more white and black thin strips in the 94\AA\ image, while the other is more uniformly colored.  

The root mean square average of the amplitude excursion obtained from this model is 
17\%  for 94\AA\  channel, and 9\% for  335\AA\  channel, not too far from those measured in the data.
As explained by \cite{guarrasi010}, the ultimate reason for this behavior is that overall each strand spends proportionally a long time, and with a high emission measure, at temperatures around 3~MK, i.e. just the temperature of maximum sensitivity of the 335\AA\ channel. Extrapolating this result to a whole active region, threaded with thousands of magnetic field lines, the loops will therefore be more populated by bright strands and appear more uniformly bright in this channel. On the contrary, they will appear less ''filled" in a hotter channel, consistently with what we observe in the 94\AA\ channel. 

Similar results are obtained with pulses that heat the plasma to slightly higher or lower temperatures ($\approx20-30\%$). Simulations and model no longer match the observations for weaker pulses leading to plasma maximum temperature $\sim4$ MK and lower, where the cool $\sim1$ MK component would dominate the observed emission in the 94\AA\ channel. We can also exclude models with heat pulses lasting significantly less than 1~minute: the main hot emitting ions would not have the time to reach ionization equilibrium \citep{reale08}, therefore effectively suppressing the high temperature emission, visible in the 94\AA\ channel. 
As a warning, we remark that this model is tailored to describe a system of some loops, rather than an entire active region, and that the center of an active region may require a different (larger) number of strands per pixel.

As a further support,  we only mention here that the model predicts the emission in the 94\AA\ channel to vary {\it in time} with larger amplitude than in the 335\AA\ channel, and that this is confirmed by the observation: the pixel light curves obtained from the model are in  good agreement with the observed ones. We defer a detailed report on time variability to forthcoming work.

\section{Conclusions}

The new SDO active region data show evidence of extensive hot plasma, which is very finely structured, as expected in a scenario
where storms of intense and rapid energy pulses are heating multi-stranded
coronal loops. The strands are temporarily heated to flare temperatures. 
Our analysis is straightforward, and little depends on processing details, such as background subtraction. 
% The results are found already from visual inspection (Fig.~\ref{fig:ar}): the image at 335\AA\  shows the plasma at 3 MK, the image in the 94\AA\  channel  shows plasma at 6 - 8 MK along with plasma at 1 MK (making identification of each not straightforward) and the image at 171\AA\  clearly identifies the plasma at 1 MK.
We remark that in the 94\AA\  channel we are not resolving the individual strands. We instead still see bundles of strands and the brightest ones are those where the fraction of very hot strands is relatively large.
The reason why these very hot components are so difficult to detect is their small emission measure, and their small duty cycle with respect to most of their evolution time, spent mostly in the subsequent long cooling phase. This result supports previous debated analyses \citep{reale09,reale09b,schmelz09,mctiernan09,sylwester010}, is consistent with many other pieces of evidence pointing to dynamically heated loops, e.g. overdensity of $\sim1$ MK loops \citep{klimchuk06,reale010}, and  supports localized heat pulses, e.g., nanoflares, to play an important role in powering coronal active regions.

\bigskip
\acknowledgements{FR, GP, MG acknowledge support from  Italian Ministero dell'Universit\`a e Ricerca and Agenzia Spaziale Italiana (ASI), contract I/023/09/0.  PT was supported by contract SP02H1701R from Lockheed-Martin to the Smithsonian Astrophysical Observatory.}

\bibliographystyle{apj}
%\bibliography{references}
% \bibliography{biblio}

% \begin{thebibliography}{12}
% \expandafter\ifx\csname natexlab\endcsname\relax\def\natexlab#1{#1}\fi
% 
% \bibitem[{{Cargill}(1995)}]{1995itsa.conf...17C}
% {Cargill}, P.~J. 1995, in Infrared tools for solar astrophysics: What's next?,
%   ed. J.~R. {Kuhn} \& M.~J. {Penn}, 17--+
% 
% 
% \end{thebibliography}
\end{document}